\RequirePackage{fix-cm}

\documentclass{Latex/aastex702}

\usepackage{bm}
\usepackage{url}
\usepackage{amsmath}
\usepackage{amssymb}
\usepackage{upgreek}
\usepackage{float}
\usepackage{graphicx}
\usepackage[caption=false]{subfig}
\usepackage{booktabs}

\newcommand{\scell}[3][c]{%
  \begin{tabular}[#1]{@{}#2@{}}#3\end{tabular}}
\newcommand{\com}[1]{}

\def\UrlBreaks{\do\/\do-}

\begin{document}

\title{\textbf{Physical Characterization of Moon Impactor 2025-010D}}

\author[0000-0001-9418-1663]{Tanner Campbell}
\affiliation{Space4 Center, 
             The University of Arizona, Tucson, AZ, USA}
\affiliation{Lunar \& Planetary Laboratory, The University of Arizona, 
             Tucson, AZ, USA}
\email{}

\author[0000-0002-4412-5732]{Adam Battle$^*$}
\affiliation{Space4 Center, 
             The University of Arizona, Tucson, AZ, USA}
\affiliation{Lunar \& Planetary Laboratory, The University of Arizona, 
             Tucson, AZ, USA}
\email{}

\author{Bill Gray}
\affiliation{Project Pluto, Bowdoinham, ME, USA}
\email{}

\author[0000-0002-0764-4672]{Juan A. Sanchez}
\affiliation{Planetary Science Institute, Tucson, AZ, USA}
\email{}

\author[0000-0001-6018-1729]{David C. Cantillo}
\affiliation{Space4 Center, 
             The University of Arizona, Tucson, AZ, USA}
\affiliation{Lunar \& Planetary Laboratory, The University of Arizona, 
             Tucson, AZ, USA}
\email{}

\author[0000-0003-0349-7932]{Lucille Le Corre}
\affiliation{Planetary Science Institute, Tucson, AZ, USA}
\email{}

\author[0000-0002-7743-3491]{Vishnu Reddy}
\affiliation{Space4 Center, 
             The University of Arizona, Tucson, AZ, USA}
\affiliation{Lunar \& Planetary Laboratory, The University of Arizona, 
             Tucson, AZ, USA}
\email{}

{\let\thefootnote\relax\footnote{* Corresponding author: adambattle@arizona.edu}}

\begin{abstract}
    Renewed interest in lunar exploration is creating a growing population of poorly tracked cislunar objects, some of which will ultimately impact the Moon. The ultimate fate of rocket upper stages or failed mission payloads is often unknown and unplanned. Determining the origin and physical properties of such objects will become increasingly important as sustained lunar exploration places humans and infrastructure on the surface. Here we present a ground-based physical and dynamical characterization of 2025-010D, a Falcon 9 upper stage predicted to impact the lunar farside on 2026 August 05. Backward propagation of its orbit independently links the object to the ``Ghost Riders in the Sky" launch, while visible and near-infrared spectroscopy distinguishes it from natural objects and reveals absorption bands consistent with spacecraft thermal control materials. Photometric lightcurves confirm its elongated shape and reveal an unexpected change of more than 8 s in its $\sim$7 min rotation period during observations. We predict an impact between Bell and Einstein craters at $\sim$2.36 km s$^{-1}$, producing a crater $\lesssim$25 m in diameter. Since its provenance is independently known, 2025-010D provides a benchmark for identifying and characterizing future cislunar objects of uncertain origin.
\end{abstract}


\section{Introduction}\label{intro}
A SpaceX Falcon 9 Block 5 second stage collided with the Moon on 2026 August 05 after only a year and a half in orbit. This Falcon 9 carried the ``Ghost Riders in the Sky" (GRS) mission with both the Firefly Aerospace Blue Ghost Mission 1 (M1) and the Japan Aerospace Exploration Agency (JAXA) Hakuto-R Mission 2 (M2) payloads \citep{CLC2025, MMD2026, TMC2026}. Both missions successfully separated from the second stage rocket body (R/B), and on 2025 March 02 Blue Ghost M1 became the first commercial spacecraft to perform a soft-landing on the Moon \citep{MMD2026}. Hakuto-R M2 suffered an anomaly with its laser range finder and crashed into the lunar surface on 2025 June 05 UTC \citep{I2025}. After separation, these spacecraft were given international designators 2025-010A and 2025-010B respectively, and a payload adapter was dubbed 2025-010C. The second stage R/B was given the international designator 2025-010D and the North American Aerospace Defense Command ID (NORAD ID) 62719. The payload adapter reentered Earth's atmosphere, but 2025-010D has been in a highly elliptical ($r_{p} \approx 9,400$ km, $r_{a} \approx 377,000$ km) geocentric orbit since launch \citep{USSC2026}.

Dozens of R/Bs and payloads have crashed into the Moon due to equipment malfunctions and occasionally, an upper stage has been intentionally rammed into the Moon to conduct science experiments \citep{WNP2017}. During the Apollo missions, several upper stages were crashed so that surface equipment could measure the resulting vibrations and scientists could learn about the Moon's interior structure \citep{LED1970}. Similarly, the Lunar Crater Observation and Sensing Satellite (LCROSS) conducted an experiment to measure water ice ejected in the resulting plume of its disposed upper stage. The LCROSS impact utilized an unusually low-density, low-velocity impactor which improved cratering models as well as our understanding of the regolith characteristics in volatile-rich, permanently shadowed craters on the Moon \citep{E2023}. In some cases, the lunar disposal has been intentional to ensure a controlled collision and avoid interference with spacecraft in orbit around the Moon or with infrastructure on the surface \citep{WNP2017, S2019}

Impact experiments on the Moon and other planetary bodies are conducted to refine crater formation models and improve our general understanding of planetary surfaces. NASA's Double Asteroid Redirect Test (DART) was a controlled impact into asteroid (65803) Didymos's moon to study the efficacy of using an impact to change an asteroid's orbit \citep{CRC2024}. The DART mission, several ground-based observations, and the follow-up Hera mission to the same system focused on the ejecta from the impact event and the change in orbit of the companion asteroid which indicates the impact efficiency \citep{MKB2022, KTL2023, CAB2023}. In crater modeling, crater diameter and depth are heavily influenced by surface properties such as subsurface porosity, terrain strength, volatile abundance, and impacted material density \citep{M1989, H1993}. Studies that leverage a priori information  of an impactor have furthered the understanding of crater formation mechanisms such as NASA's Deep Impact experiment at Comet 9P/Temple 1 \citep{ABD2005, HH2007} and the JAXA Hyabusa2 mission at (162173) Ryugu \citep{SHM2019, WHH2019, KMI2019}. Additionally, imaging orbiters around the Moon such as NASA's Lunar Reconnaissance Orbiter (LRO, \citealt{CBF2007}) and the Korea Pathfinder Lunar Orbiter (KPLO, \citealt{JCK2024}) provide opportunities for similar experiments to occur on the Moon serendipitously, via before and after imagery, when there is advanced notice that an object will crash into the Moon, either intentionally or not.

Excluding equipment failures and planned collisions, only one other R/B is known to have unintentionally impacted the Moon \citep{CBG2023}. Leading up to the historic 2022 lunar impact, the origin of WE0913A was highly debated, but a detailed astrodynamics analysis, supported by visible spectroscopy and post-impact in-situ imaging determined the object to be the Chang'e 5-T1 mission's upper stage \citep{CBG2023}. As a result, the upcoming 2025-010D impact builds on an uncomfortable precedent. With nearly 100 missions planned for the Moon in the next decade \citep{HRS2014, KPN2022, X2022, NASA2026a} compared to $\sim$35 in the last 3 \citep{PS2026, NASA2026b}, this likely will be a more frequent occurrence. Not only does the risk of impact rise with more frequent lunar missions, the risk \textit{from} impact rises as well. Planned missions include both crewed surface missions and permanent infrastructure \citep{LWW2019, KPN2022} which will be at risk from impacts by human-made space debris and natural objects. In addition to the infrastructure risk, debris impacts are, at their core, junk left on the Moon. Although disposal on the surface is sometimes unavoidable, the Outer Space Treaty of 1967 encourages the limit to pollution of celestial bodies such as the Moon \citep{UN1967} so future consideration needs to be given to the appropriate disposal of equipment during mission planning.

The US Space Force uses the Space Surveillance Network (SSN) to track most artificial satellites, but due to the vast distances and proximity to the Moon, artificial objects like 2025-010D are often poorly tracked and characterized. Planetary defense surveys scan cislunar space for natural objects and provide occasional astrometric observations of artificial objects, however, more studies are needed to mature techniques used to characterize targets in cislunar space. In this paper we present techniques to resolve ambiguities in the origin and nature of an object like 2025-010D given no a priori information, with a description of the methods given in Section \ref{meth}. We show an astrodynamic analysis of 2025-010D's orbit and determine the launch time and location (Section \ref{traj}), recovering the object's origin accurately ($<$100 km on Earth and $\sim$5 s in time). A visible and near-infrared spectral analysis is presented to determine the reflective properties of artificial objects in deep space and absorption bands are identified as well as potential material analogs based on this remote sensing data. We compare these results to common asteroid spectral characteristics to better inform future studies seeking to understand whether a cislunar object is composed of natural or artificial materials (Section \ref{spec}). Photometric lightcurves are presented to estimate the upper stage's rotation period, changes to the period over time, the size, and the axial size ratio (Section \ref{light}). Predictions of the lunar impact time, location, and crater properties are given (Section \ref{TaP}). Finally, we present a discussion of how these techniques can resolve ambiguities in the origin and nature of an object like 2025-010D. We compare the trajectory and crater morphology to post-impact imagery from lunar orbiters, and we discuss the implications of these accidental impactors (Section \ref{disc}).

\section{Methods}\label{meth}
\subsection{Observations}\label{obs}
Some astrometric measurements used for orbit determination, particularly early 
measurements used for back propagation (see Section \ref{traj}), were collected 
by planetary defense surveys and made available through the Minor Planet 
Center (MPC)\footnote[1]{https://www.minorplanetcenter.net/}. Additional astrometry, as well as all photometry and 
spectroscopy analyzed were collected using the telescope systems given in Table \ref{tb0}. 

\begin{table}[H]
    \centering
    \caption{Relevant parameters for the sensors used to collect data of the 2025-010D R/B.}
    \label{tb0}
    \begin{tabular}{lcccc}
        \toprule
        \textbf{Name} & \scell[c]{c}{\textbf{Location}} & \scell[c]{c}{\textbf{Aperture} \\ \textbf{(m)}} & \scell[c]{c}{\textbf{FOV} \\ \textbf{(arcmin)}} & \scell[c]{c}{\textbf{Pixel Size} \\ \textbf{($\frac{as}{px}$)}} \\
        \noalign{\smallskip}\hline\hline
        CONDOR & \scell[c]{c}{Arizona, USA} & \scell[c]{c}{1.0} & \scell[c]{c}{43 x 32} & \scell[c]{c}{0.18} \\
        RAPTORS I & \scell[c]{c}{Arizona, USA} & \scell[c]{c}{0.61} & \scell[c]{c}{16 x 16} & \scell[c]{c}{0.95} \\
        RAPTORS II & \scell[c]{c}{Arizona, USA} & \scell[c]{c}{0.5} & \scell[c]{c}{98 x 74} & \scell[c]{c}{0.51} \\
        IRTF & \scell[c]{c}{Hawai`i, USA} & \scell[c]{c}{3.2} & \scell[c]{c}{0.013 x 0.25} & \scell[c]{c}{0.15} \\
        iTelescope T32 & \scell[c]{c}{New South Wales, AUS} & \scell[c]{c}{0.43} & \scell[c]{c}{42 x 28} & \scell[c]{c}{0.27} \\
        iTelescope T59 & \scell[c]{c}{New South Wales, AUS} & \scell[c]{c}{0.51} & \scell[c]{c}{37 x 37} & \scell[c]{c}{0.54} \\
        \bottomrule
        \multicolumn{5}{l}{\textbf{Notes.} The IRTF FOV denotes slit dimensions.}
    \end{tabular}
\end{table}

\com{01/10: 6.9?, 01/16: 92.7, 01/31: 13.7, 02/01: 20.9568, 
     02/24: 24.4622, 03/22: 22.6981, 04/18: 24.2331, 05/12: 75.0952, 
     05/17: 42.3242, 05/20: 25.5, 05/24: 3.0733, 06/24: 40.1796}
\com{01/10: 173.1, 01/16: 87.3, 01/31: 166.3, 02/01: 159.5, 
     02/24: 155.5, 03/22: 157.3, 04/18: 155.7, 05/12: 104.8, 
     05/17: 137.8, 05/20: 154.5, 05/24: 176.9, 06/24: 139.7}

\begin{table}[H]
    \centering
    \caption{Summary of observation epoch and geometry for all of the collected data of 2025-010D.}
    \label{tb1}
    \setlength\tabcolsep{0pt}
    \begin{tabular}{lcccccc}
        \toprule
        \textbf{Name} & \scell[c]{c}{\textbf{Date} \\ \textbf{(UTC)}} & \scell[c]{c}{\textbf{Range} \\ \textbf{(km)}} & \scell[c]{c}{\textbf{V Mag} \\ \textbf{(Pred.)}} & \scell[c]{c}{\textbf{Phase Angle} \\ \textbf{($^{\circ}$)}} & \scell[c]{c}{\textbf{Obs. Type}} & \scell[c]{c}{\textbf{No. of Obs.}} \\
        \noalign{\smallskip}\hline\hline
        CONDOR & \scell[c]{c}{2026 Jan 10 \\ 2026 Feb 01 \\ 2026 Feb 24 \\ 2026 Mar 22 \\ 2026 Apr 18 \\ 2026 Jun 24} & \scell[c]{c}{187,800 \\ 261,500 \\ 412,500 \\ 465,200 \\ 507,300 \\ 559,800} & \scell[c]{c}{12.9 \\ 14.1 \\ 15.2 \\ 15.4 \\ 15.7 \\ 16.3} & \scell[c]{c}{6.9 \\ 21.0 \\ 24.5 \\ 22.7 \\ 24.2 \\ 40.2} & \scell[c]{c}{AP \\ S \\ APS \\ AP \\ AP \\ AP} & \scell[c]{c}{791 \\ 309 \\ 332 \\ 300 \\ 401 \\ 283} \\
        \hline
        RAPTORS I & \scell[c]{c}{2026 Jan 31} & \scell[c]{c}{211,800} & \scell[c]{c}{13.4} & \scell[c]{c}{13.7} & \scell[c]{c}{S} & \scell[c]{c}{93} \\
        \hline
        RAPTORS II & \scell[c]{c}{2026 May 12 \\ 2026 May 24} & \scell[c]{c}{453,000 \\ 520,500} & \scell[c]{c}{16.9 \\ 14.9} & \scell[c]{c}{75.1 \\ 3.1} & \scell[c]{c}{AP \\ AP} & \scell[c]{c}{677 \\ 333} \\
        \hline
        IRTF & \scell[c]{c}{2026 Jan 16 \\ 2026 May 17} & \scell[c]{c}{443,800 \\ 534,700} & \scell[c]{c}{17.5 \\ 16.3} & \scell[c]{c}{92.7 \\ 42.3} & \scell[c]{c}{S \\ S} & \scell[c]{c}{16 \\ 28} \\
        \hline
        iTelescope T32 & \scell[c]{c}{2026 Jul 30 \\ 2026 Aug 02 \\ 2026 Aug 04} & \scell[c]{c}{316,600 \\ 301,000 \\ 346,800} & \scell[c]{c}{15.5 \\ 15.5 \\ 16.8} & \scell[c]{c}{34.5 \\ 41.5 \\ 73.9} & \scell[c]{c}{A \\ A \\ A} & \scell[c]{c}{6 \\ 3 \\ 5} \\
        \hline
        iTelescope T59 & \scell[c]{c}{2026 Jul 29 \\ 2026 Jul 31 \\ 2026 Aug 01} & \scell[c]{c}{339,600 \\ 302,800 \\ 295,600} & \scell[c]{c}{16.0 \\ 15.1 \\ 15.1} & \scell[c]{c}{49.2 \\ 25.1 \\ 27.3} & \scell[c]{c}{A \\ AP \\ A} & \scell[c]{c}{5 \\ 200 \\ 5} \\
        \bottomrule
        \multicolumn{7}{l}{\textbf{Notes.} Observation types are (A)strometry, (P)hotometry, and/or (S)pectroscopy.}
    \end{tabular}
\end{table}

All astrometric and photometric observations collected for this study used a 
clear or luminance filter, with a mixture of both sidereal and object-rate tracking on 
different nights as well as a range of exposure times from 0.5 - 15 s 
depending on the object motion and brightness. Exposure times for visible spectroscopy ranged from 3 - 60 s and near infrared spectra used 200 s exposures. Due to the telescope plate 
scales (Table \ref{tb0}) and the typical local sight seeing conditions, data from the CONDOR telescope were collected at 4x binning and data from RAPTORS II, T32, and T59 were 2x binned. All spectrometric observations collected using the RAPTORS I 
telescope were taken with a 30 line/mm transmission diffraction grating, which 
results in a 0.0146 $\upmu$m/px spectral resolution (R$\sim$30 at 0.45 $\upmu$m; \citealt{BRS2022, BRF2023}). All spectra collected with the CONDOR telescope used a 30 line/mm transmission diffraction grating (0.0152 $\upmu$m/px spectral resolution, R$\sim$30 at 0.45 $\upmu$m). The data from the NASA Infrared Telescope Facility (IRTF) was captured with the SpeX instrument \citep{RTO2003} in prism mode (R$\sim$100). A summary of the observational circumstances for all data collected on the 2025-010D R/B is given in Table \ref{tb1}, for observational circumstances of the spectral comparison objects, please see Table \ref{tb2}.

\subsection{Material Comparison Data}\label{matel}
Laboratory spectra of AZ Technologies AZJ-4020 white thermal control material (TCM) were acquired with an ASD LabSpec 4 Hi-Res spectrometer which has a 3 nm resolution at 0.7 $\upmu$m and 6 nm resolution at 1.4 and 2.1 $\upmu$m (R$\sim$230 - 350). Spectra were measured relative to a baseline Spectralon disk and both the sample and the Spectralon were illuminated by a 120 W quartz-tungsten bulb with a 0$^{\circ}$ incidence angle and 30$^{\circ}$ reflected angle. Data were corrected for dark current, known infrared features of Spectralon, and detector sensitivity offsets at 1 and 1.8 nm \citep{KCS2017}. These methods are the same as previous laboratory studies of spacecraft materials at NIR wavelengths such as the TCM polyvinyl fluoride (PVF) that was studied in \citet{BRS2024}. Continuum-removed spectra of the AZJ-4020 TCM and the targets are shown in Section \ref{spec} to better compare their band centers.

\subsection{Astrometry and Photometry}\label{ap}
Astrometric and photometric data analysis is performed making use of 
the Gaia DR2 star catalog \citep{GAIADR2} for both astrometric plate solutions 
(to third degree and order) and photometric calibration (linear, to Gaia G band) \citep{CFR2022, CBG2023}. Across the five nights with astrometric/photometric data from the CONDOR telescope the aggregate photometric calibration RMS is 0.098 magnitudes. For the two nights from the RAPTORS II telescope the aggregate photometric calibration RMS is 0.020 magnitudes. The one night from the T59 telescope that had photometric data (2026 Jul 31) had a photometric RMS of 0.042 magnitudes. The astrometric plate solutions are calculated with a maximum allowable star position fit RMS of 0.33 arcseconds.

The reduced magnitude is calculated with a distance modulus that includes the Sun-target distance and Earth-target distance, as described in \citet{B2010}. The average of each night's lightcurve is used for the brightness value to remove amplitude effects from the object's periodic rotation. The H-G model from \citet{BHD1989} is fit to the data (Equations \ref{eq2}-\ref{eq7}), weighted by the measurement uncertainty via least squares minimization (Section \ref{light}).

\begin{equation}
    \label{eq2}
    H(\alpha) = H_{0}-2.5{\log}_{10}[(1-G)\Phi_{1}(\alpha)+G\Phi_{2}(\alpha)]
\end{equation}

\begin{equation}
    \label{eq3}
    \Phi_{i}(\alpha) = W\Phi_{i,S}(\alpha)+(1-W)\Phi_{i,L}(\alpha); \quad i \in \{1,2\}
\end{equation}

\begin{equation}
    \label{eq4}
    W = e^{-90.56\tan^{2}(\frac{\alpha}{2})}
\end{equation}

\begin{equation}
    \label{eq5}
    \Phi_{i,S}(\alpha) = 1-C_{i}\frac{\sin\alpha}{0.119+1.341\sin\alpha-0.754\sin^{2}\alpha}
\end{equation}

\begin{equation}
    \label{eq6}
    \Phi_{i,L}(\alpha) = e^{-A_{i}\tan^{B_{i}}(\frac{\alpha}{2})}
\end{equation}

\begin{equation}
    \label{eq7}
    A = \{3.332, 1.862\}; B = \{0.631, 1.218\}; C = \{0.986, 0.238\}
\end{equation}

\subsection{Fourier Analysis}\label{four}
The Fourier analysis is done by considering pairs of time and magnitude data 
at each observation epoch and fitting a function of the form in Equation 
\ref{eq1} via robust nonlinear least-squares weighted by the measurement 
uncertainty.

\begin{equation}
    \label{eq1}
    f(t) = a_{0}+a_{1}\cos(wt)+b_{1}\sin(wt)+a_{2}\cos(2wt)+b_{2}\sin(2wt)
\end{equation}

The estimated lightcurve period (with confidence bounds) is obtained directly 
from the estimated parameters, and the amplitude is calculated from the 
resultant minimum error Fourier series. This method is sensitive to gaps in 
the data, and to data that is sampled uniformly in a nearly harmonic cadence 
with the period of the lightcurve. In either case the Lomb-Scargle periodogram 
may be more useful to extract a fundamental period, though it lacks 
flexibility in how data can be modeled \citep{L1976, S1982}.

\subsection{Visible Spectroscopy}\label{vis}
Slitless visible spectra (0.45 – 0.82 $\upmu$m) were collected with the RAPTORS I and CONDOR telescopes. Images were processed using standard CCD techniques of dark and bias frame subtraction. A solar analog star was observed each night to produce reflectance measurements and standard stars were observed most nights to correct for airmass effects (see Table \ref{tb2}) when the difference between target and solar analog airmasses exceeded 0.1. 

\begin{table}[H]
    \centering
    \caption{Observational circumstances for spectroscopy data collection including solar analog and standard star names used for data processing.}
    \label{tb2}
    \setlength\tabcolsep{0pt}
    \begin{tabular}{lcccccc}
        \toprule
        \textbf{Name} & \scell[c]{c}{\textbf{Date} \\ \textbf{(UTC)}} & \scell[c]{c}{\textbf{Object}} & \scell[c]{c}{\textbf{Airmass}} & \scell[c]{c}{\textbf{Standard Star}} & \scell[c]{c}{\textbf{Solar Analog}} & \scell[c]{c}{\textbf{Wavelength} \\ \textbf{($\upmu$m)}} \\
        \noalign{\smallskip}\hline\hline
        CONDOR & \scell[c]{c}{2026 Feb 01 \\ 2026 Feb 24} & \scell[c]{c}{2025-010D \\ 2025-010D} & \scell[c]{c}{1.06 - 1.52 \\ 1.00 - 1.13} & \scell[c]{c}{HD 88961 \\ Not Used} & \scell[c]{c}{SAO 109542 \\ SAO 120107} & \scell[c]{c}{0.456 - 0.851} \\
        \hline
        RAPTORS I & \scell[c]{c}{2026 Jan 31} & \scell[c]{c}{2025-010D} & \scell[c]{c}{1.00 - 1.23} & \scell[c]{c}{HD 76765} & \scell[c]{c}{SAO 109542} & \scell[c]{c}{0.453 - 0.847} \\
        \hline
        IRTF & \scell[c]{c}{2026 Jan 16 \\ 2026 May 17 \\ 2026 May 20} & \scell[c]{c}{2025-010D \\ 2025-010D \\ 2024-127B} & \scell[c]{c}{1.08 - 1.23 \\ 1.02 - 1.30 \\ 1.08} & \scell[c]{c}{HD 123049 \\ SAO 100585 \\ HD 111662} & \scell[c]{c}{SAO 120107} & \scell[c]{c}{0.696 - 2.499 \\ 0.688 - 2.499 \\ 0.691 - 2.500} \\
        \bottomrule
    \end{tabular}
\end{table}

Incident light from a point source generates both a zeroth order point source (straight path through the grating) and typically only the first order spectrum due to the use of blazed (0.405 $\upmu$m) transmission diffraction gratings. Counts are summed along the column of each pixel within a user-defined extraction box and the corresponding wavelength is calculated by the number of columns from the centroid of the zeroth-order point source. A background estimation box is used to reduce effects of sky background on the spectrum. This process is repeated for each target image to produce a median spectrum with error bars represented by the standard deviation. The same process is performed for standard stars and solar analogs which are applied to the main target. When the solar analog is close in airmass to the target, reflectance is given by simply dividing the target spectrum by the solar analog spectrum. When a standard star is needed to correct for airmass effects, the target and solar analog spectra are both independently divided by the standard star spectrum and then the resulting airmass-corrected measurements are divided by each other as before \citep{BRS2022, BRF2023}. Visible spectra presented in this study are normalized at 0.73 $\upmu$m to provide overlap with NIR data and are then renormalized at 1.5 $\upmu$m for comparison with relevant planetary materials.

One unique challenge for collecting spectroscopy of elongated, tumbling objects such as rocket bodies is the large change in apparent magnitude, as shown in Section \ref{light}. Because spectroscopy spreads the light from the target across multiple pixels, there is a large reduction in sensitivity compared to the same sensor being used for astrometry or photometry. Due to the periodic 0.52 - 2.16 mag change in brightness of 2025-010D, only the brightest $\sim$50\% of visible spectra observed could be used while still obtaining a good signal of the target. A preliminary comparison of the fainter spectra on a given night was performed and no obvious difference existed between the fainter and brighter visible spectra, but significant noise was introduced across all wavelengths. The bias toward the brighter spectra means a bias toward observing the long side (higher surface area) of the target and that particular combination of materials. 

\subsection{Color-Index Estimation}\label{color}
Typically, color indices derived from broadband filter photometry are used to estimate a four-color spectrum using Equation \ref{eq8}, which can be derived directly from the standard magnitude equation \citep{P1856}. The reflectance in one filter, R($\lambda_1$) chosen as the normalization point and is set to equal one at that filter's effective wavelength. The reflectance in a second filter, R($\lambda_2$), is estimated at the effective wavelength of the second filter's band pass using Equation \ref{eq8}. The color index in these two filters is given by $m_1$ - $m_2$ and the corresponding solar color index in the same two filters is given by $m_{1,\odot} - m_{2,\odot}$. The standard is to make $m_1$ the bluer of the two filters used for the color index (e.g., g-r). 

\begin{equation}
    \label{eq8}
    \frac{R(\lambda_{2})}{R(\lambda_{1})} = 10^{0.4[(m_{1}-m_{2}) - (m_{1,\odot} - m_{2,\odot})]}
\end{equation}

A representative color index in any photometric system can be approximated from our visible spectroscopy data. The Sloan Digital Sky Survey (SDSS, \citealt{JSR2005, DTF2010}) and Johnson-Cousins \citep{JM1953, C1976} photometric systems are chosen for their wide use in astronomical research and for comparisons to asteroid values. Equation \ref{eq9} rearranges Equation \ref{eq8} to solve for the representative color index given a visible spectrum of the object. 

\begin{equation}
    \label{eq9}
    m_{1}-m_{2} =  (m_{1,\odot}-m_{2,\odot}) + 2.5\log_{10}\left(\frac{R(\lambda_{2})}{R(\lambda_{1})}\right)
\end{equation}

For the SDSS filters, the ``2.5m Reference" effective wavelengths from \citet{DTF2010} are used: $\lambda_g$ = 0.4627 $\upmu$m, $\lambda_r$ = 0.6140 $\upmu$m, $\lambda_i$ = 0.7467 $\upmu$m, and $\lambda_z$ = 0.8887 $\upmu$m. For the Johnson-Cousins filters, the effective wavelengths reported in \citet{B2005} are used: $\lambda_B$ = 0.4361 $\upmu$m, $\lambda_V$ = 0.5448 $\upmu$m, $\lambda_R$ = 0.6407 $\upmu$m, and $\lambda_I$ = 0.7980 $\upmu$m. Solar color indices in both photometric systems are taken from \citet{HFP2006}. Because our spectra do not have data at exactly the effective wavelength of each filter, a line is fit through each night's spectrum to interpolate and determine the reflectance for each filter. The root mean square error of the fit to the spectrum is added in quadrature with the literature value's uncertainty in the solar color index to calculate the uncertainty in each color index estimate. The uncertainty in each color index estimate is on the order of 0.02 magnitudes for all bands. Color indices are estimated for each of the three nights of visible spectra and the average and standard deviation between the three nights are presented (instead of the single-measurement uncertainty) to show variation in the color index estimate.

\subsection{Near-Infrared Spectroscopy}\label{nir}
The SpeX instrument \citep{RTO2003} on NASA's Infrared Telescope Facility (IRTF) was used in prism mode to collect low resolution (R$\sim$100) near-infrared (0.7 – 2.5 $\upmu$m) spectra of 2025-010D and the comparison object, 2024-127B. The IRTF is a 3.2 m telescope located on Maunakea, Hawai`i at an elevation of approximately 4.2 km above sea level, which drastically reduces the amount of atmosphere and water vapor affecting the NIR spectra. The slit dimensions are given in Table \ref{tb0}, and observations were performed with the slit oriented along the parallactic angle to minimize differential atmospheric refraction. Data were collected by dithering the slit between A and B positions within the field of view and acquiring images in an A-B-B-A pattern to allow the subtraction of sky background between frames. To correct for telluric absorption bands, a G-type standard star was observed every hour, as well as immediately before and after each target. Observations of a solar analog are used to calculate reflectance values and to correct the spectral slope of the standard star. Due to the sensitivity of the IRTF, and the use of consecutive 200 s integration times across multiple rotations of 2025-010D, the NIR spectra should be roughly representative of a full integration of light from all sides of the object.

\subsection{Band Measurements}\label{bands}
Spectral absorption bands can be quantified by the position of the band center, band depth, and band area (see Section \ref{spec} for more details). These band parameters are measured by initially calculating a linear fit across two manually-selected peaks on either side of the band and that section of the spectrum is divided by this linear fit to flatten and normalize the band \citep{CGJ1986, STR2020}. A cubic polynomial is fit to the bottom third of the band to measure the band center and the percentage drop between the linear fit and this band center is calculated as the band depth \citep{CR1984}. Uncertainties are estimated for these band parameters by measuring the value 10 times and calculating the standard deviation \citep{STR2020}.

\subsection{Orbit Estimation}\label{od}
Orbit estimation is done by fitting one of various tunable dynamical models 
to the observed data in the least-squares sense. Observations are weighted by 
either observer-reported uncertainties or statistical analysis. Observation 
de-weighting is performed for high cadence observations (those taken for 
lightcurve analysis) to avoid artificially biasing the solution. Once an 
initial solution is fit to the best-fitting data, iterative residual threshold 
filtering is performed on all observations considered to include or exclude 
additional measurements in the solution. For back propagation and impact 
prediction, perturbations are included from variable degree and order gravity 
models for the Earth and Moon, as well as point mass contributions from the 
remaining seven planets and Pluto. By default, the orbit is computed with the Sun 
as the primary gravitational force. Nongravitational effects can be included with 
up to three additional parameters estimated ($A_{1-3}$).

\section{Trajectory Analysis}\label{traj}
To confirm the link between the lunar impactor and the Falcon 9 second stage, we conducted a trajectory analysis by using publicly available information on the launch circumstances and back propagating the trajectory calculated using optical astrometry. The GRS Falcon 9 rocket launched from Kennedy Space Center Launch Complex 39-A (LC-39A) at 06:11:39 UTC on 2025 January 15. Per the United States Space Command, it was the only rocket launch (globally) to occur that day \citep{USSC2026}. The Falcon user's guide \citep{F92025} gives two sample flight timelines but notes that every 
flight profile will differ. One example is for Low Earth Orbit (LEO), 
the other is for Geosynchronous Transfer Orbit (GTO), but neither of these is 
what we would expect from the GRS launch. In this case, each mission spent a 
prolonged period orbiting the Earth before heading to the Moon 
\citep{CLC2025, MMD2026, TMC2026}. Both profiles in the user's guide, however, 
have three burn sequences (one main stage and two second stage) followed by a 
five-minute coast before payload separation. From Firefly Aerospace's 
live mission updates page \citep{BGM2025}, launch vehicle separation is noted 
at 07:17 UTC. Assuming a similar five-minute coast, and matching the 
minute-level precision on the update website, we expect the final second stage burn 
to have completed at 07:12 UTC, marking the approximate time of first perigee for 
the R/B.

\begin{table}[H]
    \centering
    \caption{Estimated nongravitational and orbital parameters for 
             the 2025-010D R/B giving the predicted Earth periapsis 
             constraining the launch (when propagated backwards). Elements 
             are given in the Earth centered MEME J2000 reference frame.}
    \label{tb3}
    \begin{tabular}{lcc}
        \toprule
        \textbf{Parameter} & \textbf{Value} & \textbf{1$\sigma$ Uncertainty} \\
        \noalign{\smallskip}\hline\hline
        Epoch (TT) & 2025 Jan 15 08:24:00 & N/A \\
        $a$ (km) &  186501.1939 & 3.07E-2 \\
        $e$ & 0.964896381 & 1.86E-7 \\
        $i$ (deg) & 35.186047 & 8.0E-5 \\
        $\omega$ (deg) & 345.251403 & 4.5E-5 \\
        $\Omega$ (deg) & 0.106451 & 2.6E-5 \\
        $M$ (deg) & 1.93772 & 1.7E-4 \\
        $AMR$ (m$^{2}$kg$^{-1}$) & 0.0119510 & 6.74E-5 \\
        \bottomrule
    \end{tabular}
\end{table}

Using 304 observations of 2025-010D spanning 2025 January 20 to 2025 April 27 
from planetary defense surveys, made available by the Minor Planet Center 
(MPC), we can fit an orbit. We fit orbits with both 3-parameter \citep{MSY1973} 
and 1-parameter \citep{HUG2012} nongravitational effects modeled. Both were 
successful, as the baseline and quality of observations are both good enough 
to get a solution, however the uncertainties in the 3-parameter case are of 
competing magnitude to the estimated values. Using the 1-parameter 
(radial solar radiation pressure) model gives more manageable uncertainties 
and a 0.56 arcsecond mean residual to the observations. Using the estimated 
orbit and propagating backwards gives a time of perigee of 07:12:05.56 UTC 
$\pm$ 0.38 s on 2025 January 15 at 1$\upsigma$ confidence. This is 4 minutes 
and 54.44 seconds before the reported Blue Ghost M1 spacecraft separation 
\citep{BGM2025} and 5.56 seconds after the assumed first perigee/final second 
stage burn. Table \ref{tb3} gives the estimated orbital 
elements with uncertainties (Earth-centered Mean Equator Mean Equinox (MEME) 
of J2000) which constrain the launch.

\begin{figure}[H]
    \centering
    \includegraphics[width=0.8\linewidth]{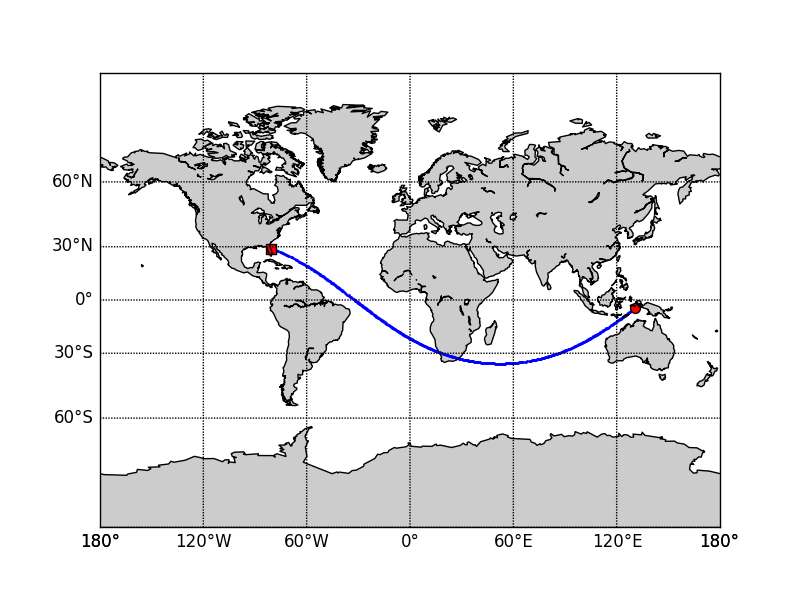}
    \caption{Ground-track (blue curve) of a circular orbit generated by taking 
             the estimate state at perigee (red dot) and setting $e = 0$ 
             and $a = r_{p}$, then propagating backwards. The LC-39A launch 
             site, from which the Blue Ghost mission 1 was launched, is shown 
             as a red square.}
    \label{figCBP}
\end{figure}

The Falcon user's guide does not give spatial trajectory information for their 
sample cases, only timing of flight events. Thus, lining up the 
backwards propagated trajectory with the sequence of launch events beyond the 
temporal analysis presented is difficult. However, with the simplifying 
assumption that the initial trajectory is along the desired orbital plane 
(plane changes are expensive and not typically done at periapsis), then it is 
possible to get an approximation of the launch path. Without estimating the full 
outbound hyperbola and engine burns, the simplest approach is to ``circularize" 
the orbit at perigee (i.e. set $e = 0$ and $a = r_{p}$) and propagate backwards 
further. A different trajectory than that of the actual launch, however the 
ground-track of this circularized orbit gives us an estimate of the flight 
corridor that the rocket traversed during the launch. Figure \ref{figCBP} shows a 
section of this circular orbit propagated backwards from the aforementioned perigee 
(large red dot), which lines up with the LC-39A launch site (large 
red square) to within 0.59 degrees ($\sim$66 km).

\section{Spectral Analysis}\label{spec}
Since we are able to determine the origin of the GRS Falcon 9 R/B, we can perform additional analyses to establish benchmark characteristics for these objects. Visible reflectance spectra have been used to create groups with similar reflectance properties, such as asteroid taxons \citep{T1984} and geostationary satellite bus types \citep{BRS2024}. Our goal here is to confirm the results of the trajectory analysis and ensure that the lunar impactor is a R/B and not a natural object such as a temporarily captured minimoon \citep{KFM2025,JBB2018}. Visible spectra are also used to monitor how surfaces change with time to investigate space aging effects \citep{PWK2020}. Analysis of absorption band center, depth, and area can be used to identify mineral chemistry, organics, water, and other materials via remote sensing \citep{L1978, K2012, HLL2021, CSL2024}. Three primary absorption bands in visible and near-infrared (NIR) wavelengths help identify planetary materials: a 0.7 $\upmu$m phyllosilicate hydration band, a $\sim$0.9 – 1.0 $\upmu$m absorption due to Fe$^{2+}$ in olivine and Ca$^{2+}$ in pyroxene, and a $\sim$2.0 $\upmu$m band due to Fe$^{2+}$ in pyroxene \citep{A1974, CGJ1986, VG1989, GBB1993}. Ultimately, many planetary materials are comprised of well-characterized silicates and spectrally similar to one another. Artificial materials, however, make up a wider range of material types including thermal control material (TCM), metals, solar panels, etc. and no comprehensive spectral database of these materials exists. We present visible (0.45 to 0.82 $\upmu$m) and NIR (0.7 to 2.5 $\upmu$m) spectral observations of 2025-010D and select comparison objects to establish a baseline at these wavelengths to aid in future identification of objects in cislunar space.

\begin{figure}[H]
    \centering
    \includegraphics[width=0.8\linewidth]{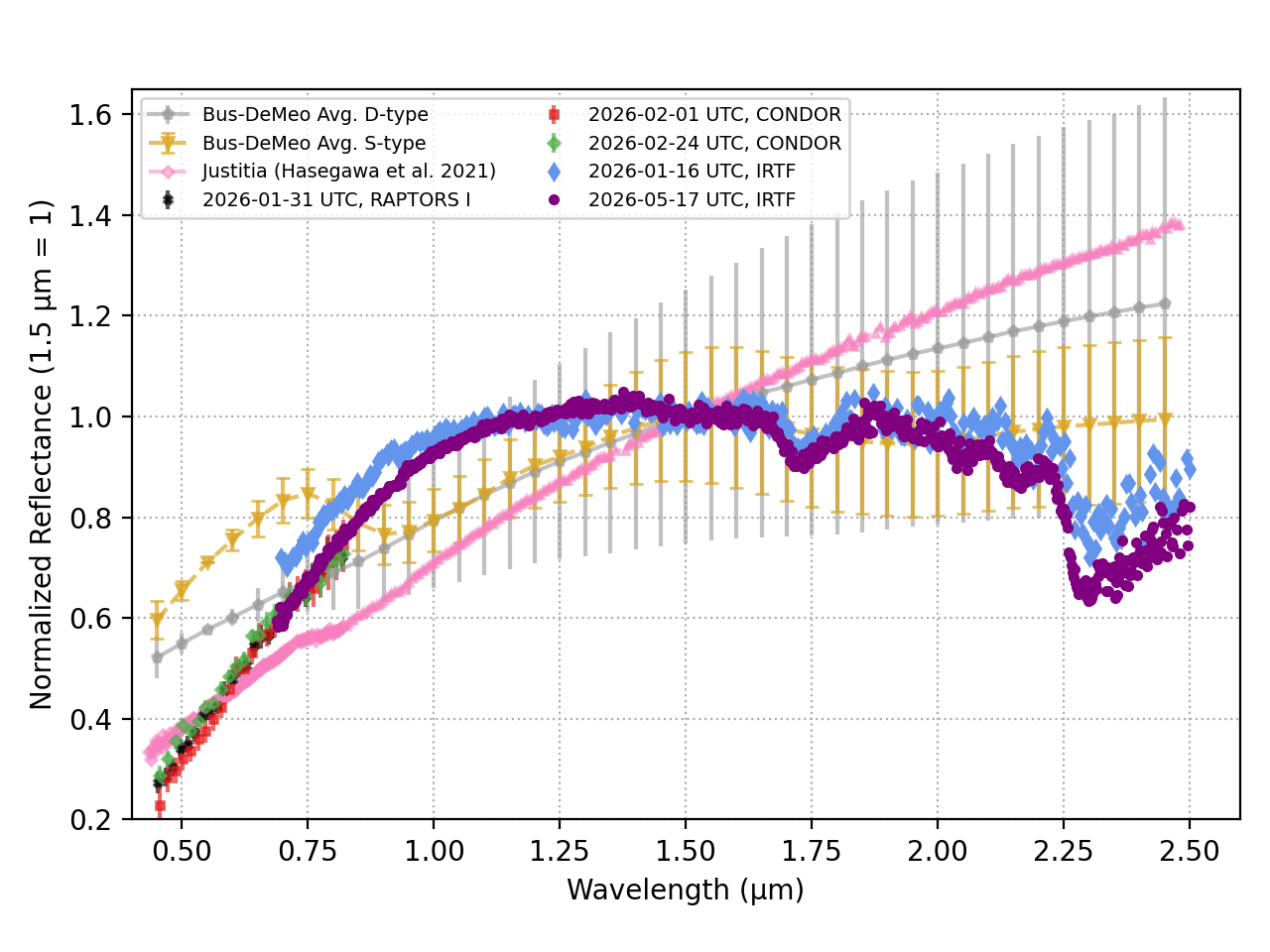}
    \caption{Visible and near infrared spectra of 2025-010D compared with the mean spectra of S- and D-type asteroids, and the spectrum of (269) Justitia from \citet{HMD2021} 2025-010D shows distinct absorption bands at $\sim$1.73 and 2.3 $\upmu$m. All spectra are normalized at 1.5 $\upmu$m with visible spectra  first being normalized at 0.73 $\upmu$m and combined with the 2026 May 17 IRTF in the overlap region before renormalizing to 1.5 $\upmu$m. Variations on the NIR spectra are due to differences in phase angle on different nights (see Table \ref{tb1}).}
    \label{figAsteroids}
\end{figure}

Figure \ref{figAsteroids} shows combined visible and NIR spectra for 2025-010D. The visible spectra are relatively featureless and have a steep, red slope with increasing reflectance at longer wavelengths. At NIR wavelengths, the spectra have a slightly blue slope and two prominent absorption bands near 1.7 and 2.3 $\upmu$m. The spectra are shown with the average spectra of S- and D-type asteroids along with bars showing their typical variation \citep{DBS2009}. Silicate-rich S-type asteroids are the most common type in near-Earth space, with 1 and 2 $\upmu$m absorption bands that differ from the prominent bands in 2025-010D's spectrum. Asteroids from the D-type taxon represent the most steeply red-slopped asteroid known in near-Earth space. A spectrum of main-belt asteroid (269) Justitia is shown, as one of the reddest planetary bodies, with a redder slope than Trojan asteroids and comparable to trans-Neptunian objects \citep{HMD2021}. Spectral slope is not diagnostic of materials due to non-compositional alteration effects such as phase angle and thermal alteration (e.g., \citealt{SRN2012, SRL2025}). Despite this, it is useful to see that 2025-010D has a redder visible spectral slope than all but the most extreme planetary material presented in the literature, as shown in Figure \ref{figAsteroids}. This rules out the likelihood of an object with these spectral properties being a natural near-Moon object such as a temporarily captured minimoon. Since visible spectroscopy may not be available for fainter targets of future studies, we approximate several color indices from our visible spectroscopy, as described in Section \ref{color}. We estimate a SDSS g-r color index of 1.02 $\pm$ 0.06, an r-i color index of 0.45 $\pm$ 0.02, and an i-z color index of 0.31 $\pm$ 0.02. Using the visible spectra, we estimate a Johnson-Cousins B-V color index of 1.13 $\pm$ 0.06, a V-R color index of 0.65 $\pm$ 0.03, and an R-I color index of 0.69 $\pm$ 0.02.

\begin{figure}[H]
    \centering
    \subfloat[Normalized Falcon 9 NIR Spectra]{%
        \includegraphics[width=0.45\linewidth]{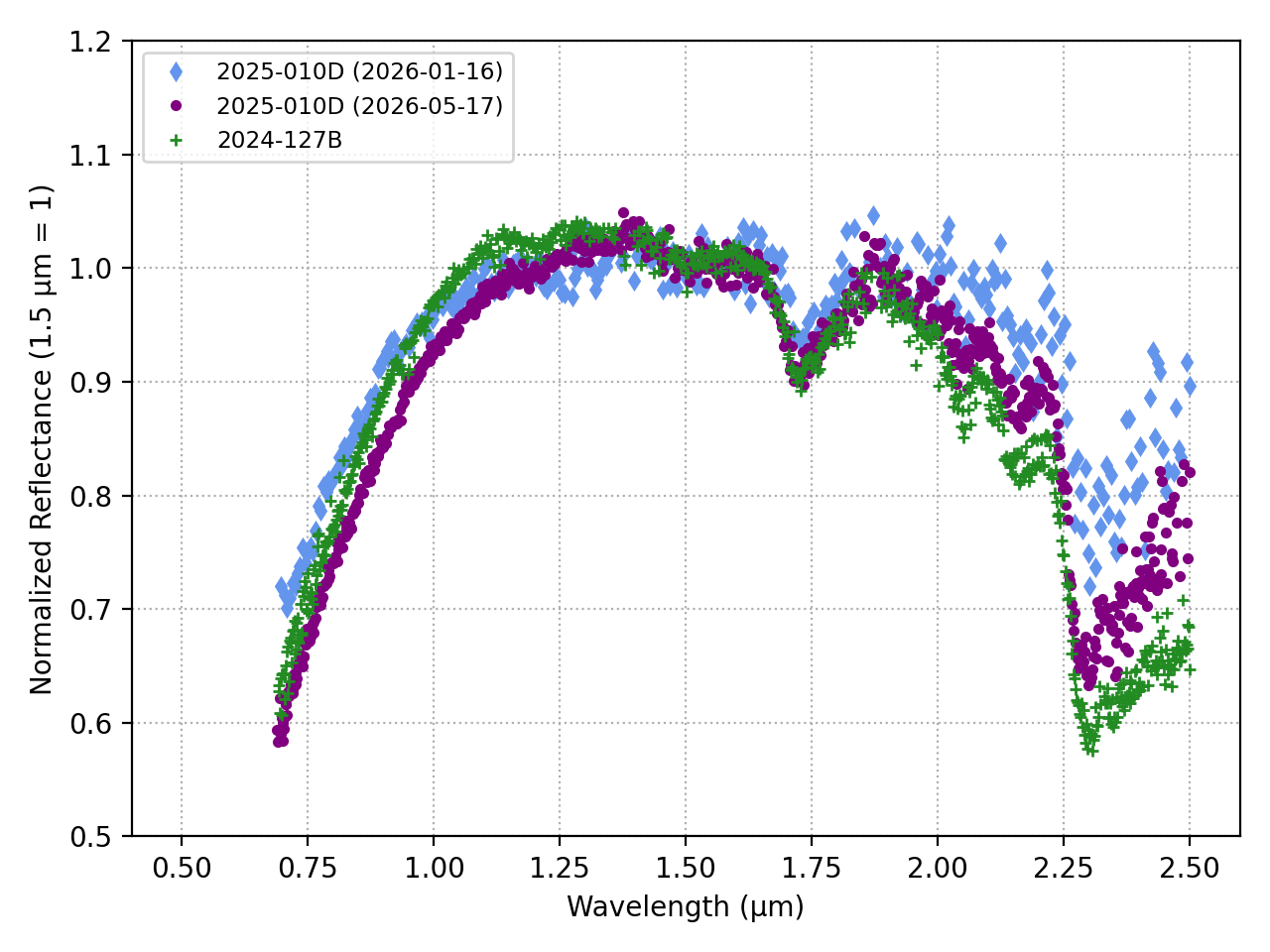}%
        \label{figF9s}%
    }
    \subfloat[Continuum-removed Spectra with AZJ-4020 TCM]{%
        \includegraphics[width=0.45\linewidth]{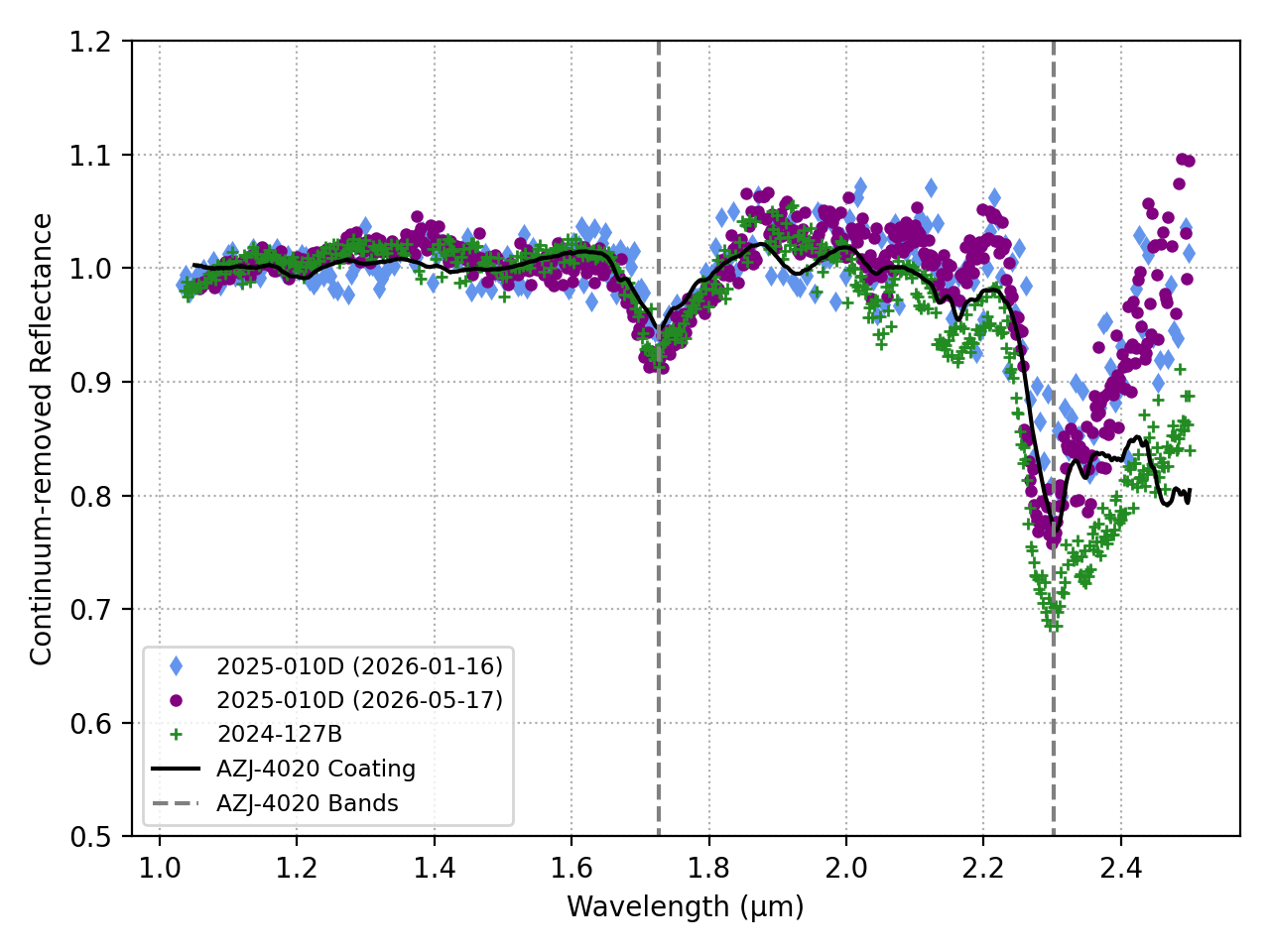}%
        \label{figAZJ}%
    }
    \caption{(a) NIR spectra of 2025-010D compared to the spectrum of another Falcon 9 upper stage, 2024-127B. Spectra are normalized at 1.5 $\upmu$m and both objects show absorption bands at $\sim$1.73 and 2.3 $\upmu$m. Variations between 2025-010D spectra are due to differences in phase angle on different nights (see Table \ref{tb1}). (b) NIR spectra of 2025-010D and 2024-127B with AZ Technology's AZJ-4020 thermal control material. Each spectrum's continuum is removed by dividing out a quadratic fit between 1 and 2 $\upmu$m. The Falcon 9 R/B absorption bands can be visually compared to those of the AZJ-4020 coating and dashed lines are shown indicating the position of the two strongest band centers for AZJ-4020. Measured band centers are shown in Table \ref{tb5}.}
    \label{figContinuum}
\end{figure}

Since spectral slope is not diagnostic, comparison of NIR absorption bands to known materials is used for direct evidence of material composition. 2025-010D is nearly identical to another Falcon 9 second stage (2024-127B) that was observed for comparison and Figure \ref{figF9s} shows the  comparison between these objects. Figure \ref{figAZJ} shows these spectra with their continua removed by fitting and then dividing each spectrum by a quadratic polynomial between approximately 1 and 2 $\upmu$m. The same continuum removal was performed for laboratory measurements of AZ Technology's AZJ-4020 \citep{AZ2015} white TCM, which was selected from an internal library of $\sim$15 spacecraft TCMs. Although this coating is a close match to what is observed on the Falcon 9 second stages, it is not necessarily what SpaceX uses and is presented here solely for comparison. 

Band centers were measured for the most prominent absorption bands in each spectrum and are listed in Table \ref{tb5}. Where possible, the band centers were measured for all Falcon 9 spectra as well as the AZJ-4020 TCM, however some bands are not discernible in the lower signal-to-noise ratio (SNR) telescope spectra and were not measured. The 1.7 $\upmu$m (Band 2) center for 2024-127B and 2025-010D matches AZJ-4020 within 3$\sigma$ and the 2.3 $\upmu$m (Band 6) center matches AZJ-4020 within 4$\sigma$ across all spectra. Additional matching absorption bands are visible in the spectrum of 2024-127B that are less obvious in spectra of 2025-010D due to the lower SNR from being a fainter object. An absorption band doublet (Band 5) is noted for AZJ-4020 that is not resolved in telescopic data. The AZJ-4020 coating absorption bands are likely due to organic compounds from the epoxy-based coating \citep{GCB2012, RBG2022}. Although we cannot say that the observed material is specifically the AZJ-4020 TCM, the true material likely has a similar composition to produce absorption bands so similar to this TCM.

\begin{table}[H]
    \centering
    \caption{Band centers measured from IRTF NIR observations of the targets in this study and laboratory samples of an analogous thermal control material: AZ Technology's AZJ-4020 coating.}
    \label{tb5}
    \begin{tabular}{lcccccc}
        \toprule
        \textbf{Name} & \textbf{Band 1} & \textbf{Band 2} & \textbf{Band 3} & \textbf{Band 4} & \textbf{Band 5} & \textbf{Band 6} \\
        \noalign{\smallskip}\hline\hline
        2025-010D$^{a}$ & -- & 1.739 $\pm$ 0.005 & -- & -- & -- & 2.314 $\pm$ 0.006 \\
        2025-010D$^{b}$ &  -- & 1.725 $\pm$ 0.001 & -- & -- & 2.162 $\pm$ 0.004 & 2.292 $\pm$ 0.003 \\
        2024-127B & 1.214 $\pm$ 0.010 & 1.724 $\pm$ 0.003 & -- & 2.048 $\pm$ 0.002 & 2.142 $\pm$ 0.004 & 2.294 $\pm$ 0.003 \\
        AZJ-4020 & 1.212 $\pm$ 0.002 & 1.727 $\pm$ 0.001 & 1.928 $\pm$ 0.001 & 2.042 $\pm$ 0.001 & 2.135 \& 2.163 $\pm$ 0.001 & 2.304 $\pm$ 0.001 \\
        \bottomrule
        \multicolumn{7}{l}{\textbf{Notes.} Band centers are in $\upmu$m. Band 6 for AZJ 4020 is a doublet.} \\
        \multicolumn{7}{l}{$^{a}$2026 Jan 16} \\
        \multicolumn{7}{l}{$^{b}$2026 May 17}
    \end{tabular}
\end{table}

\section{Photometric Analysis}\label{light}
In addition to the surface properties that can be determined by spectroscopy, 
photometry can help determine other physical and dynamical properties of the observed 
target such as size, spin state, or even anomalous events. 
All photometric measurements taken of 2025-010D from 2026 January 
through July exhibit consistent periodicity. By fitting a two-parameter 
Fourier series to the photometric measurements at each epoch via nonlinear 
least-squares, we can estimate the period and the amplitude of the 
lightcurve which are in turn indicative of the spin period and axial ratio 
of the object \citep{KLL1992a, KLL1992b, MTW2020, KBO2010, KPL2010, CFR2022, 
CBG2023}. The Fourier series amplitudes are 0.52 - 2.16 magnitudes across the 
data set. This amplitude range does not represent specular glints seen from 
the object and only encompasses the underlying structure of the lightcurve 
which is dominated by the diffusive reflectance of the body \citep{CFR2022}. 
Using the relationship for lightcurve amplitude and phase angle to axial 
ratio \citep{KBO2010} gives 3.17:1 as an estimated maximum lower bound in axial ratio. 
The Falcon user's guide \citep{F92025} gives a second stage diameter of 3.66 m, 
however no length is given. Estimates range from 10 m to 16 m (axial ratio 2.73:1 
to 4.37:1) depending on the source \citep{B2015, SLR2017}, with 13.8 m (3.77:1) 
being common \citep{P2026, W2026}. The Falcon 9 uses the same second stage as the 
Falcon Heavy, with one estimate \citep{S2022} being 12.6 m (axial ratio 3.44:1). The 
relationship for axial ratio is empirically derived from 
asteroid observations \citep{KBO2010}. However, by using the Fourier series 
(thereby ignoring the specular glints), the phase-angle-amplitude-lightcurve relationship of a 
R/B becomes similar to that of an ellipsoidal asteroid \citep{GDO2006, KBO2010, 
KPL2010, CFR2022, CBG2023}, and indeed in this case we see the estimated axial 
ratio lower bound is appropriate.

\begin{figure}[H]
    \centering
    \subfloat[CONDOR telescope, 2026 Jan 10]{%
        \includegraphics[width=0.33\linewidth]{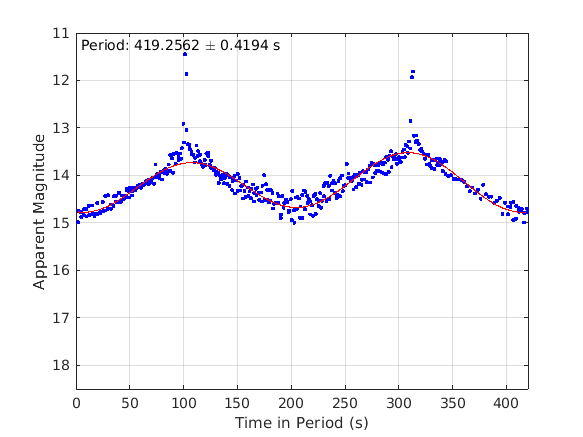}%
        \label{figL01}%
    }
    \subfloat[CONDOR telescope, 2026 Feb 24]{%
        \includegraphics[width=0.33\linewidth]{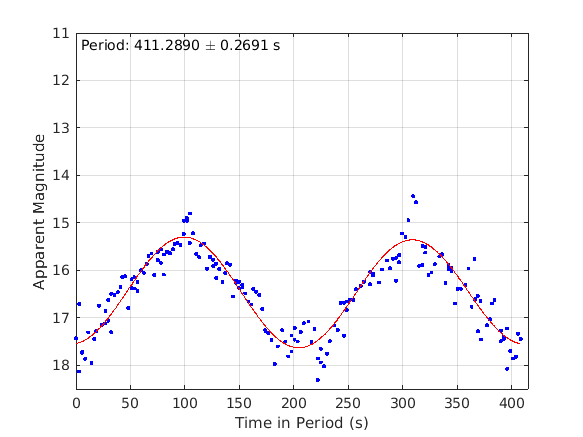}%
        \label{figL02}%
    }
    \subfloat[CONDOR telescope, 2026 Mar 22]{%
        \includegraphics[width=0.33\linewidth]{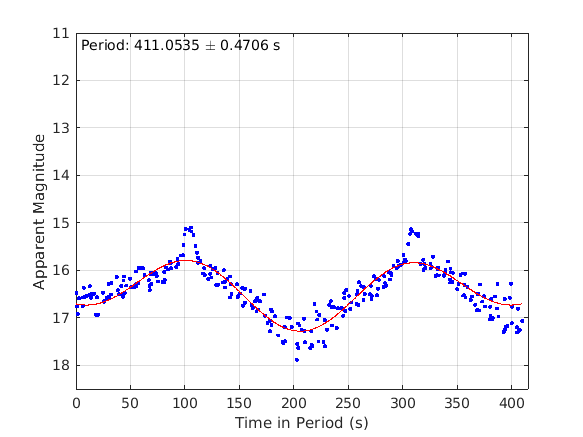}%
        \label{figL03}%
    } \\
    \subfloat[CONDOR telescope, 2026 Apr 18]{%
        \includegraphics[width=0.33\linewidth]{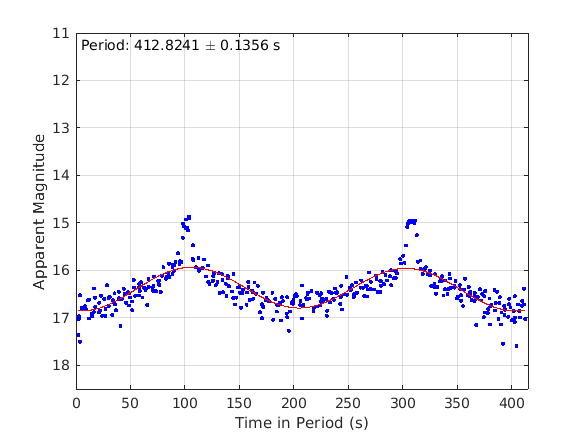}%
        \label{figL04}%
    }
    \subfloat[RAPTORS II telescope, 2026 May 12]{%
        \includegraphics[width=0.33\linewidth]{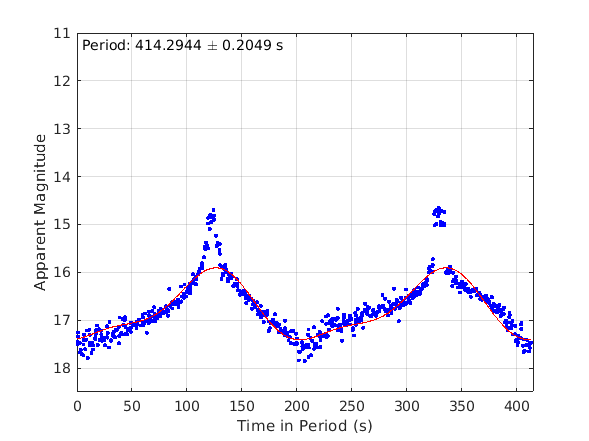}%
        \label{figL05}%
    }
    \subfloat[RAPTORS II telescope, 2026 May 24]{%
        \includegraphics[width=0.33\linewidth]{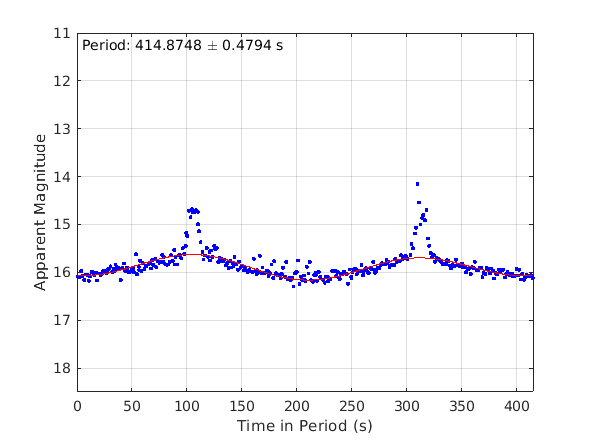}%
        \label{figL06}%
    } \\
    \subfloat[CONDOR telescope, 2026 Jun 24]{%
        \includegraphics[width=0.33\linewidth]{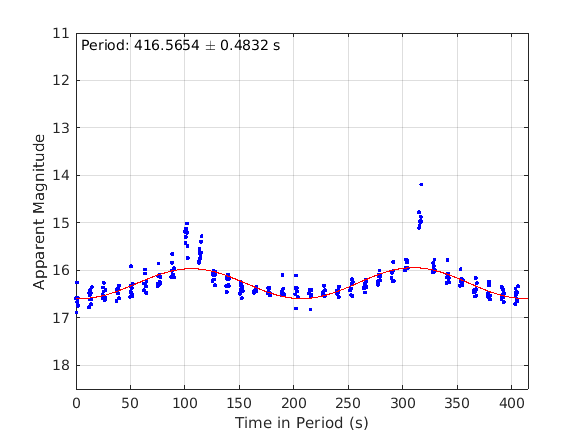}%
        \label{figL07}%
    }
        \subfloat[T59 telescope, 2026 Jul 31]{%
        \includegraphics[width=0.33\linewidth]{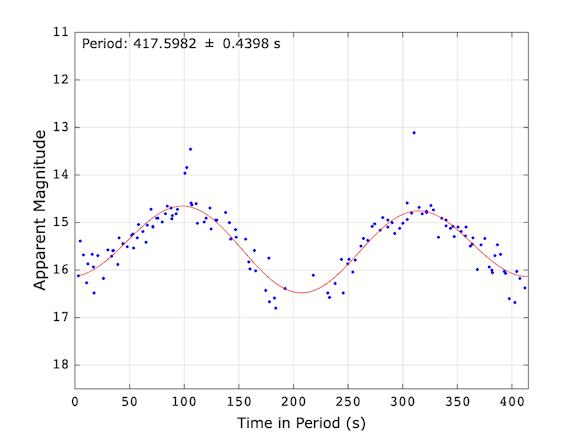}%
        \label{figL08}%
    }
    \caption{Lightcurves of the 2025-010D R/B from the CONDOR, 
             RAPTORS II, and T59 telescopes. The lightcurves have been period-wrapped by 
             fitting a two parameter Fourier series (red curves) to the data 
             (blue dots) via nonlinear least-squares minimization.} 
    \label{figLC}
\end{figure}

\begin{figure}[H]
    \centering
    \includegraphics[width=0.65\linewidth]{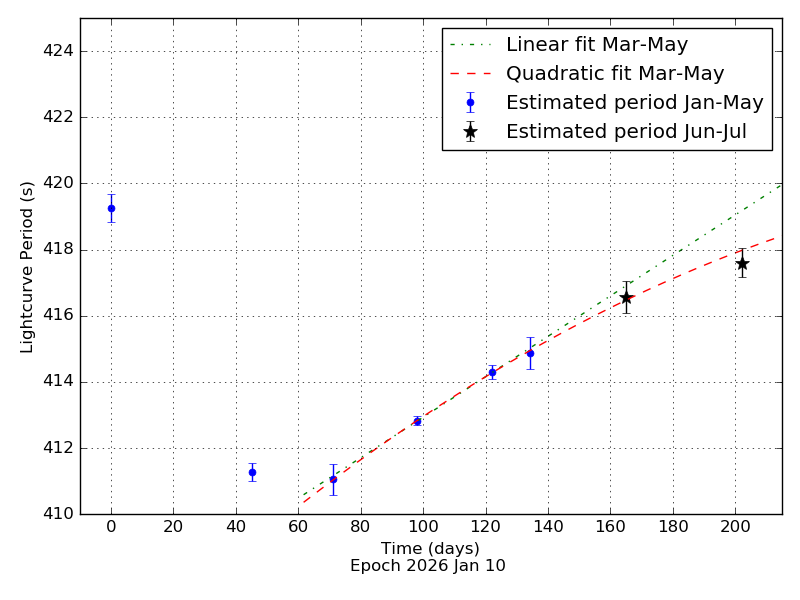}
    \caption{The estimated lightcurve periods from January through May (blue 
             dots) show significant variation with time. Both a linear (green 
             dot-dash) and quadratic (red dash) model were fit through the 
             measurements from March through May and used as a predictor for 
             the lightcurve period in the 2026 June 24 and July 31 data (black stars). The linear prediction matches the Fourier fit period of the 2026 June 
             24 data to 0.34 s, and the quadratic prediction matches to 0.09 s, 
             both within the Fourier estimate uncertainty. For the 2026 July 31 data, the linear fit is outside of the 3$\upsigma$ fit uncertainty, but the quadratic fit is still within 1$\upsigma$ with an error of 0.38 s.}
    \label{figPDF}
\end{figure}

The period estimates from the Fourier analysis range from 411.0535 $\pm$ 
0.4706 s to 419.2562 $\pm$ 0.4194 s. The period-wrapped lightcurves from 
all observations are shown in Figure \ref{figLC}. Surprisingly, 
the lightcurve period decreases by approximately 8 seconds ($\sim$19$\sigma$) between 
2026 January 10 and 2026 February 24 (see Figure \ref{figPDF}). The estimated period on 2026 Mar 22 is 
slightly faster in mean value than that on February 24, however both estimates fall within 1$\upsigma$ 
of each other and have a low Hellinger distance ($\sim$0.33) so we can not distinguish them from two 
characterizations of the same period \citep{H1909}. This decrease in lightcurve period 
(indicative of spin-up of the R/B) is unexpected; the natural rotational dynamics of a defunct rocket in space 
are expected to be almost entirely dissipative \citep{W2025}. This adds to the difficulty in ascribing the 
observed lightcurve period decrease to a specific cause. We do not know what happened between the observation 
epochs in January and February, so we cannot say if there was an `instantaneous' decrease in the 
period, or if it was gradual. Moreover, we do not know if the period continued to decrease between February 
and March to some lower minimum, or if it was stable throughout. Gravity, drag, 
magnetic fields, internal body flexure, left over fuel, and solar radiation 
all cause dampening of an object's rotation. Solar radiation, excess fuel 
outgassing, or debris impact each could cause an increase in 
rotation rate. Solar radiation appears in both categories as it can exhibit 
either effect depending on the object's surface properties and orientation 
\citep{BCC1969}. A collision (debris or otherwise) would cause a near instantaneous spin-up of the R/B, 
causing a sudden decrease in the lightcurve period. Fuel outgassing could be released from a buildup of 
pressure in a large spurt, or slowly over time. And similarly solar radiation would typically slowly modify 
the lightcurve period \citep{ASM2015, ASC2018}, however a large outburst from the Sun (solar flare) cause a 
short-term change to the R/B rotation. More photometric observations before January and between 
January and March would be needed in order to be able to characterize the nature of the spin-up, however 
regardless of the cause, under the generalized dissipative assumption there will 
be exponential decay into a minimum energy `flat spin' state after this event \citep{W2025}. 

In Figure \ref{figPDF}, we can see that after the `spin-up', the 
lightcurve period decreases through the remaining observations. Due to the 
uncertainty in the estimated periods, and the resultant divergence in exponential 
estimation, both linear and quadratic functions have been fit to 
the lightcurve periods from 2026 March 22 through 2026 May 24 instead. As a 
validation of the method for short-term prediction (barring the ability to 
estimate the exponential decay), the fits excluded the data from 2026 June 24 
and July 31 and were used as an estimator for comparison (see Figure \ref{figPDF}). On June 24, the linear fit yields a period of 416.908 s and the quadratic fit gives 416.472 s, both within 1$\upsigma$ of the Fourier estimate (416.5654 $\pm$ 0.4832, Figure \ref{figL07}) at that epoch. The quadratic fit matches to 0.093 s, however this method of estimation will break down as the estimation window grows and the quadratic estimator deviates from the underlying exponential decay. On July 31, the linear fit gives a period of 419.19 s and is outside of the 3$\upsigma$ uncertainty for that Fourier estimate. The quadratic fit, however, is still within 1$\upsigma$ of the Fourier estimate giving 417.983 s (a 0.38 s difference).

Since quadratics have well-defined extrema, we can estimate a maximum period 
of 420.374 s based on our data, which is within 3$\upsigma$ of the lightcurve 
period on 2026 January 10. If 2025-010D was near to its minimum energy 
rotational state in January, we would expect the rotation to decay to 
approximately the same level. However, since we do not have photometric 
measurements appropriate for estimating the lightcurve period from before 
2026 January, we cannot know what the rotational state was.

\begin{figure}[H]
    \centering
    \includegraphics[width=0.65\linewidth]{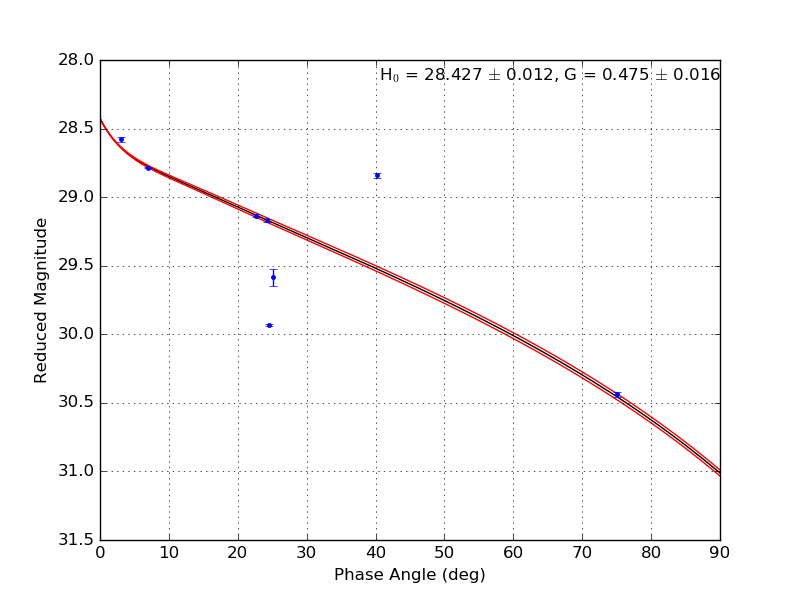}
    \caption{Phase curve of the 2025-010D R/B using photometric data from the CONDOR, RAPTORS II, and T59 telescopes (blue dots). The data from 25 - 40$^{\circ}$ phase angle is excluded in the fit since this behavior is not represented by the standard H-G model \citep{B2010}. The best-fit curve is shown in black and the 1$\upsigma$ variation in G with fixed H is shown in red on either side of the best fit.}
    \label{figHG}
\end{figure}

The photometric observations for this study span a range of phase angles from 3.1 - 75.1$^{\circ}$ (Table \ref{tb1}) which, allows us to estimate the absolute magnitude (H$_{0}$) and slope parameter (G) \citep{BHD1989, B2010} for 2025-010D (Equation \ref{eq2}). The reduced magnitude is calculated as described in Section \ref{ap} and phase angles are retrieved from the orbital data. The average brightness of each lightcurve is used to remove the effects of amplitude variation on the data. We exclude the data between 25 - 40$^{\circ}$ phase angle when fitting the H-G model, as this behavior is not accounted for in the phase functions and is likely an artifact of the different reflective dynamics of artificial objects compared to the natural objects for which this model was derived, but it is impossible to say without more data. The fit results in an absolute magnitude in the Gaia G photometric system of H$_{\mathrm{G}}$ = 28.427 $\pm$ 0.012 and a slope parameter of G = 0.475 $\pm$ 0.016, as shown in Figure \ref{figHG}. Uncertainty in the estimation of H correlates to changes in the y-intercept of the fit, but the uncertainty in G is less intuitive where variance converges towards lower phase angles, and diverges from the best-fit at larger phase angles. Figure \ref{figHG} also shows the 1$\upsigma$ bounds of varying G with H held constant to illustrate the effect. Typically, phase curves are more densely populated, having a more comprehensive sampling in phase angle space. Additionally, it is important (exponentially so) to have densely sampled data in the low phase angle regions to have an accurate H-G model of the object. The low uncertainty on both H and G is an artifact of the small number of data points and the relative measurement uncertainties so these results should be taken as preliminary with more data needed to have confidence in the model. A better sampling of phase angles could also give a better understanding of what is occurring in the 25 - 40$^{\circ}$ phase angle range.

\section{Lunar Impact Predictions}\label{TaP}
\begin{figure}[H]
    \centering
    \includegraphics[width=0.8\linewidth]{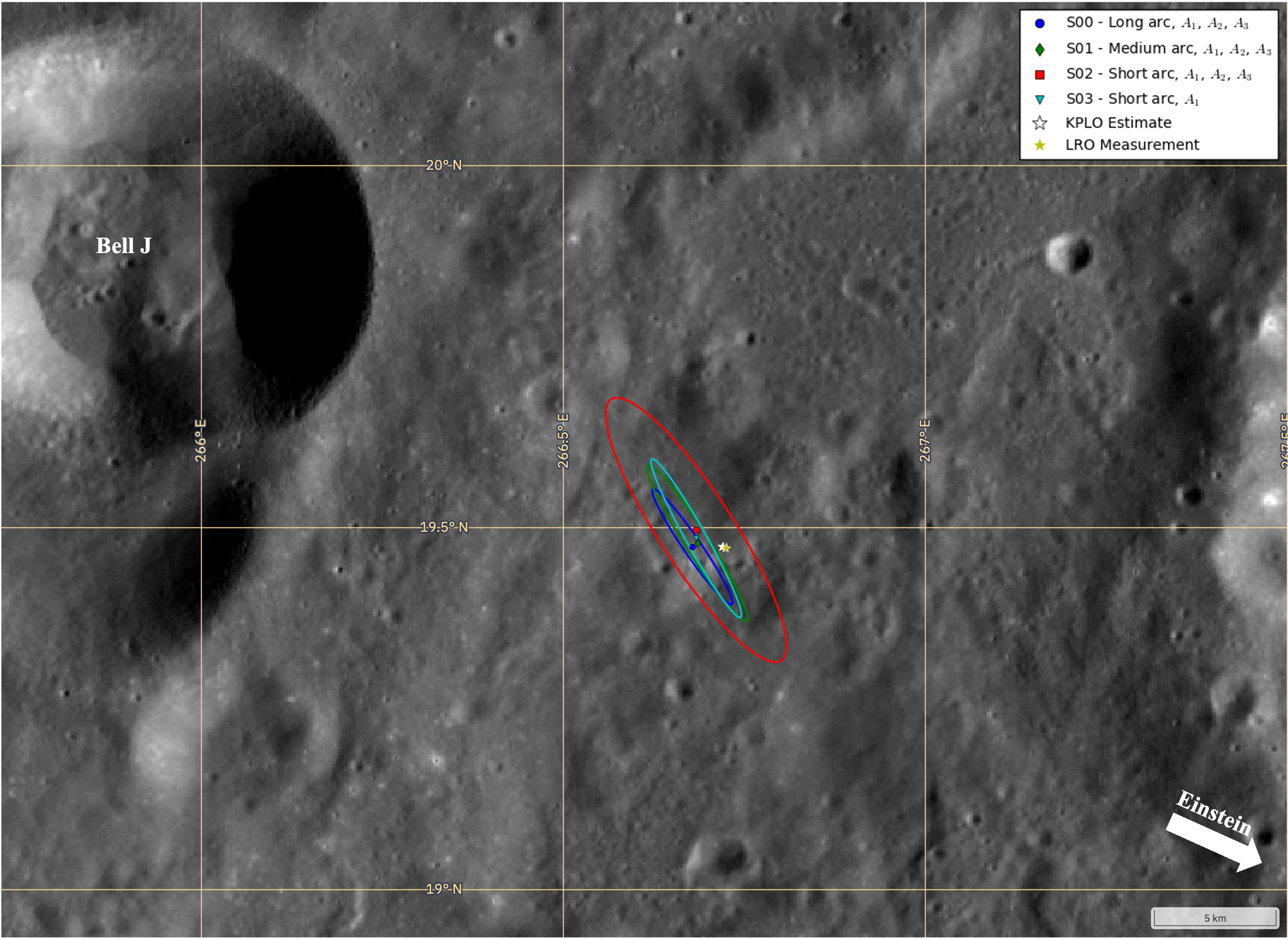}
    \caption{Impact location prediction for four different trajectory 
             solutions including the 3$\upsigma$ uncertainties in the 
             predictions. Each uncertainty region is generated via 1000 
             Monte Carlo propagations, sampled from the trajectory solution 
             covariance. Background is from Lunar QuickMap 
             (https://quickmap.lroc.asu.edu). The Korea Pathfinder Lunar Orbiter (KPLO) position (white star) was estimated from the images released by The Korea AeroSpace Administration (KASA) post impact \citep{KASA2026}. The Lunar Reconnaissance Orbiter (LRO) position (yellow star) was published by the LRO Space Operations Center \citep{LRO2026}.}
    \label{figIPR}
\end{figure}

To estimate the lunar impact location, four different orbit 
solutions were computed that fit the observations. The solutions varied 
in number of observations included and number of nongravitational 
parameters estimated. The shortest arc (S02 and S03) included 260 measurements from 
2026 May 21 to 2026 Aug 04, and the longest (S00) used 443 observations from 
2026 February 24 to 2026 Aug 04. All four solutions fit the data 
with mean residuals less than 0.5 arcseconds. After determining 
the orbits, each covariance was sampled 1000 times and propagated forward. 
All 4000 trajectories impact the Moon, and the distributions of the impact 
locations are given in Figure \ref{figIPR}. All four solutions agree in mean impact 
location to within 0.025 degrees. We propose the short arc which includes 3 nongravitational parameters (S02) as the solution given in Table \ref{tb4} since all other simulated impacts are within the 3$\upsigma$ uncertainty of this solution. 

\begin{table}[H]
    \centering
    \caption{Estimated nongravitational and osculating orbital parameters for 
             the 2025-010D R/B giving the predicted lunar impact. Elements are 
             given in the Earth centered MEME J2000 reference frame.}
    \label{tb4}
    \begin{tabular}{lcc}
        \toprule
        \textbf{Parameter} & \textbf{Value} & \textbf{1$\sigma$ Uncertainty} \\
        \noalign{\smallskip}\hline\hline
        Epoch (TT) & 2026 Aug 05 04:48:00 & N/A \\
        $a$ (km) & 1901079.52 & 468.24 \\
        $e$ & 0.8619589 & 3.39E-5 \\
        $i$ (deg) & 65.4919 & 0.0014 \\
        $\omega$ (deg) & 308.9190 & 0.0007 \\
        $\Omega$ (deg) & 20.77841 & 0.00044 \\
        $M$ (deg) & 3.2818 & 0.0012 \\
        $A_{1}$ (km s$^{-2}$) & 4.7835E-11 & 1.072E-12 \\
        $A_{2}$ (km s$^{-2}$) & -1.604E-12 & 1.275E-12 \\
        $A_{3}$ (km s$^{-2}$) & -5.756E-12 & 8.26E-13 \\
        \bottomrule
    \end{tabular}
\end{table}

We find a mean impact time of 2026 August 05 06:35:36.026 UTC $\pm$ 5.110 s at 
latitude 19.497 $\pm$ 0.061$^{\circ}$ and longitude 266.677 $\pm$ 0.017$^{\circ}$ 
between the Bell and Einstein craters, just beyond the northwest limb on the far side
of the Moon. The 3-parameter model does a good job of estimating nongravitational effects on an orbit, but it assumes that the scale of the 
perturbations (the $A_{1-3}$ parameters) are fixed in time. This assumption is 
typically not valid for very long arcs and limits the accuracy of future predictions, which contributes to our choice of S02 as the proposed solution.

From the orbital state given in Table \ref{tb4}, we estimate 
an angle just over 32$^{\circ}$ above the local horizon at $\sim$2.36 km s$^{-1}$ 
for the impact. Crater ellipticity is nearly constant (and $\sim$1) for impact 
angles above 20$^{\circ}$, so we expect the crater 
to be nearly circular \citep{MHM2017}. Using depth-to-diameter ratios ($d/D$) \citep{HRH2024}, we would expect $d/D \lesssim$ 0.1, and from energy-diameter scaling \citep{H1993, M1989, WHH2003}, the 
diameter should be $\lesssim$25 m. Impact crater morphology is very 
difficult to predict, and can vary widely based on local site density and 
material characteristics \citep{MHM2017, H1993, M1989}.

\section{Discussion}\label{disc}
The techniques presented in this research demonstrate the ability to eliminate ambiguities in the origin or nature of cislunar objects. A detailed astrodynamic analysis links the object 2025-010D (NORAD ID 62719) to the launch of Blue Ghost M1 and Hakuto-R M2 to within 66 km and 5.56 s. We predict a mean impact time of 2026 August 05 06:35:36.026 UTC $\pm$ 5.110 s at latitude 19.497 $\pm$ 0.061$^{\circ}$ and longitude 266.677 $\pm$ 0.017$^{\circ}$. Before an official impact location had been released, post-impact images were released by the Korea AeroSpace Administration (KASA) and we estimate our solution to be within $\sim$1 km of the feature identified in those images \citep{KASA2026}. NASA's Lunar Reconnaissance Orbiter (LRO) later imaged the impact crater and determined the impact site to be at 19.476$^{\circ}$ N 266.714$^{\circ}$ E, a difference of 0.04$^{\circ}$ ($\sim$1.2 km) from our mean estimate \citep{LRO2026}. This falls just outside the the 3$\upsigma$ uncertainty ellipses of all but the S02 solution, which highlights the difficulty and care that must be taken when estimating nongravitational effects, particularly of artificial objects. Standard 1-3 parameter models that are meant for natural objects fit well for short time periods, but the complex surface properties of artificial objects mean these estimations can break down over longer periods.

Using a fully nadir impact angle, 3$\upsigma$ upper bound impact speed, generous mass estimates for the R/B, and simplifying assumptions for the subsurface properties \citep{MHM2017, H1993, M1989}, we give an upper bound in crater diameter of 25 m. Using mean values, we estimate $\sim$17 m. Based on the study conducted in \citet{HRH2024}, impacts such as this should have a depth-to-diameter ratio of approximately 0.1. LRO was able to image the crater sufficiently to estimate the diameter at 18 m and the depth at $<$3 m ($d/D <$ 0.16), agreeing well with our predictions. LRO was able to take images with multiple viewing angles, which yields some information on the subsurface regolith that was exposed by the impact \citep{LRO2026}, and given further analysis this data will could help to constrain future crater modeling of other similar impacts. 

Photometric analysis shows that the object had a lightcurve amplitude suggestive of a lower bound in axial ratio of 3.17:1 which is consistent with the shape of R/Bs and published estimates for the size of a Falcon 9 second stage \citep{F92025, B2015, SLR2017, P2026, W2026, S2022}. The lightcurves changed in period over the course of the observation campaign in an unexpected manner. The period was originally estimated at 419.256 $\pm{0.419}$ s, and rapidly increased to 411.054 $\pm{0.471}$ s. This increase is difficult to explain, but could be due to excess fuel outgassing, collision (e.g., debris), or solar radiation. This spin-up was followed by a decay in period almost back to the original state. This rotational decay is expected for an object free-rotating in a dissipative system, which should follow an exponential. We show the utility of using first and second degree polynomials in predicting short-term rotation rates during this observed decay, predicting observed lightcurve periods to within 0.5 s. The quadratic allows the estimation of a maximum rotation period (minimum energy state), for which we find $\sim$420.37 s, which is very close to the initially observed lightcurve period. The spin-up and decay of the lightcurve period is unlikely to be caused just by Yarkovsky–O'Keefe–Radzievskii–Paddack (YORP) effect even though it has been applied to the study of defunct satellites \citep{ASM2015, ASC2018}. YORP effect is a very small torque, and therefore it requires long timescales to cause appreciable effect (order of years), and does not torque symmetrical (both geometrically and compositionally) bodies. The spin-up observed occurred in a less-than-a-month time scale, and R/Bs are mostly symmetrical. The root cause of the spin-up is likely never to be known, where more observations surrounding the period decrease would be necessary to accurately characterize the event. 

Using average photometric measurements from each night we estimate an absolute magnitude, H$_{0}$, and slope parameter, G, for the target \citep{BHD1989, B2010}. We find an absolute magnitude in the Gaia G band of 28.427 $\pm$ 0.012. Using our estimate of the target's B-V color of 1.13, this translates to a typical absolute magnitude in the V band of H$_{\mathrm{V}}$ = 28.96 $\pm$ 0.38, where the larger uncertainty is a result of the magnitude system transformation \citep{JGC2010}. The H$_{\mathrm{V}}$ and 1$\upsigma$ range suggest an object with a spherical-equivalent diameter of 1.9 - 8.1 m assuming an albedo from 0.9 - 0.1 \citep{BHD1989, H1997}. Artificial objects are comprised of materials with varied albedos, so picking one representative albedo is always challenging, but we note that an albedo of 0.344 results in a diameter estimate of 3.66 m (the short-axis diameter of the Falcon 9 second stage) \citep{F92025}. The slope parameter, G, is determined to be 0.475 $\pm$ 0.016 which is unaffected by the magnitude system transformation and would be notably high if it were a natural body. The average and standard deviation of the slope parameters of 583 asteroids measured by \citet{PHK2012} is 0.21 $\pm$ 0.09, with only 25 asteroids having slope parameters higher than 0.4. Our estimate of G for 2025-010D is near the upper 3$\upsigma$ boundary for natural bodies, and an unusually high G value may be a useful identifier of artificial objects in future studies. We note that our phase curve in Figure \ref{figHG} is sparsely populated and more data would improve the estimates of H and G, which may be sensitive to under sampling in phase angle space, particularly at phase angles $\lesssim$10$^{\circ}$.

Spectroscopic analysis showed a surface more red-slopped in visible wavelengths than all types of near-Earth asteroids and ultra-red Main Belt asteroids such as (269) Justitia. Using photometry derived from our visible spectra, we estimate an SDSS g-r color of 1.02 $\pm$ 0.06 and a Johnson-Cousins B-V color of 1.13 $\pm$ 0.06 with additional color indices given in Section \ref{spec}. The g-r color estimate is much larger (more red) than the average g-r color index of 0.62 for the Red Trojan population \citep{WBE2014}. Spectral slopes and color measurements are susceptible to non-compositional alteration effects, but the detection of absorption bands is more diagnostic for remote sensing spectroscopy of space object surfaces. Distinct near infrared absorption bands were identified at 1.73 and 2.3 $\upmu$m with additional bands seen in the high-SNR spectrum of comparison Falcon 9, 2024-127B. These absorption bands are consistent with organic bands seen in epoxy-based thermal control materials such as AZ Technology's AZJ-4020 coating which was used as a laboratory analog. Although we cannot say that the upper stage is coated with AZJ-4020 specifically and there is no public documentation on the paints used for the Falcon 9 upper stage, the true material likely has a similar composition such as an organic-based epoxy binder. It is worth noting that the AZJ-4020 white TCM was uniquely similar (out of $\sim$15 black and white TCMs in our internal database) to the observed NIR spectra. The observed visible and NIR spectroscopic characteristics in this study are strong benchmarks for differentiating between natural objects (e.g., minimoons and quasi-satellites) and artificial satellites that are detected by planetary defense surveys in cislunar space. 

There have been a few attempts to observe the dust and debris plume caused by the impact. The European Southern Observatory has posted a preliminary spectroscopic detection of Lithium in the dust plume and the Institute of Astrophysics of Andalusia's shared a video of plume ejecta evolution \citep{ESO2026, M2026}. Additional observations were planned by the Lunar Impact Flash Network \citep{FHG2026} and the ``Impact Flash!" citizen-science project \citep{NASA2026c} to search for a visible flash generated by the impact energy; their results are forthcoming.

There are currently no humans on the Moon, but that will change in the near future with NASA's Artemis program \citep{KPN2022} and China/Russia's plans for a lunar base \citep{LWW2019, X2022}. 2025-010D did not pose a risk to any humans or infrastructure, however the chances for harm will increase as more missions are sent to the Moon. The Outer Space Treaty of 1967 \citep{UN1967} requires all signatories to minimize harmful contamination on the Moon and other planetary systems, requiring that future thought be given to the disposal of R/Bs as we increase space traffic. Disposal of mission equipment via collision with the Moon may sometimes be unavoidable, however, so one proposal is the establishment of dedicated disposal zones, thereby limiting risk to future infrastructure and minimizing the area on the Moon that is contaminated \citep{BB2025}. 

In addition to the risk posed by human-made space debris, natural objects also pose real risk to future infrastructure on the Moon. During the short Artemis II mission, the astronauts witnessed six meteorite impacts \citep{G2026}. The Moon is bombarded by natural meteoroids every day with rates on the order of tens of impacts per hour \citep{LBX2020, LBX2024}. These meteoroids are estimated to be only hundreds of grams in mass, but they impact the Moon at $\sim$20 km $s^{-1}$ resulting in energies typically on the order of 10 - 100 MJ. The energy released by these meteoroids is comparable to the energy of fireworks and results in craters typically $<$3 m in diameter. Although 2025-010D is moving slower (2.36 km $s^{-1}$), the larger mass results in a more energetic impact ($\sim$11 GJ) and a predicted crater size of $\lesssim$25 m. Future structures on the Moon that are sufficiently shielded to withstand the background flux of meteoroids would not be sufficiently shielded to withstand an impact from an object such as 2025-010D. Natural impacts of GJ-energies are predicted to occur on the order of $<$10 times per year on the lunar surface \citep{ON1989}. 

Furthermore, the Moon is a shared human experience \citep{VLP2020, TVD2024} and should be treated with respect. For many Indigenous Peoples, the Moon is culturally and spiritually significant where some traditions view celestial bodies through relationships of kinship rather than as inert objects \citep{N2024, H2024}. In this context, human activities in space carry responsibilities to the Moon, to the Land, and to future generations. Part of this responsibility is to consider the lasting consequences of activities that leave human-made material on the lunar surface (e.g., \citet{WEEB2026}). In one well-documented example, the president of the Navajo Nation issued a critical press release after NASA's 1998 Lunar Prospector mission took human remains to the Moon \citep{H2024}. The issue resurfaced leading up to the planned 2024 Astrobotic mission carrying dozens of cremated remains where there was once again no consultation with Indigenous Nations. In this instance, the Navajo Nation president was joined by the Coalition of Large Tribes in renouncing the planned landing of cremated remains on the Moon \citep{H2024, COLT2024}. Although not specifically carrying cremated remains, rocket bodies WE0913A \citep{CBG2023} and 2025-010D are the only two known unplanned impacts (ignoring those caused by equipment malfunctions) into the Moon. Given the low number of lunar-proximity missions, the occurrence of both impacts in the span of four and a half years sets an alarming precedent for the future of lunar exploration. 

\newpage

\section*{Data Availability}
\noindent Planetary defense data was accessed through the Minor Planet Center (https://www.minorplanetcenter.net). Other data collected for this research can be made available upon reasonable request to the authors, pending academic obligations. 

\section*{Code Availability}
\noindent Astrometry and photometry can be reproduced using MPO Canopus (https://minplanobs.org). Infrared spectroscopy can be reproduced with Spextool (https://irtfweb.ifa.hawaii.edu). Orbit estimation can be reproduced with publicly available software such as OrbFit (https://adams.dm.unipi.it/orbfit) or Find\_Orb (https://www.projectpluto.com/find\_orb.htm).

\section*{Acknowledgments}
\noindent This work is supported by the state of Arizona Technology Research Initiative Fund (TRIF) grant (PI: Prof. Vishnu Reddy). This research work was supported in part by NASA Yearly Opportunities for Research in Planetary Defense Grant 80NSSC22K0514 (PI: Vishnu Reddy). Some observations reported here were obtained at the Infrared Telescope Facility, which is operated by the University of Hawai`i under Cooperative Agreement NCC 5-538 with the National Aeronautics and Space Administration, Science Mission Directorate, Planetary Astronomy Program. Some astrometric measurements used in this research came from planetary defense surveys, accessible through the Minor Planet Center (https://www.minorplanetcenter.net). We acknowledge our use of Lunar QuickMap (https://quickmap.lroc.asu.edu), a collaboration between NASA’s Lunar Reconnaissance Orbiter, Arizona State University (ASU), and Applied Coherent Technology (ACT). Most of the data in the Lunar QuickMap can be accessed at the file level from the Planetary Data System (PDS).

\newpage

\bibliographystyle{Latex/aasjournalv7.1}
\bibliography{references}

\end{document}